\documentclass[%
 aip,
 sd,%
 amsmath,amssymb,
 reprint,%
]{revtex4-1}

\usepackage{graphicx}% Include figure files
\usepackage{dcolumn}% Align table columns on decimal point
\usepackage{bm}% bold math
\usepackage{epstopdf}
\DeclareMathOperator{\sgn}{sgn}

\begin{document}

%\preprint{AIP/123-QED}

\title{Time-periodic stiffness modulation in elastic metamaterials for selective wave filtering: theory and experimental investigations}

\author{Giuseppe Trainiti}\email{gtrainiti@gatech.edu.}
\affiliation{ 
Daniel Guggenheim School of Aerospace Engineering, Georgia Institute of Technology, Atlanta, 30332, USA
}
\author{Yiwei Xia}
\affiliation{ 
George W. Woodruff School of Mechanical Engineering, Georgia Institute of Technology, Atlanta, 30332, USA
}
\author{Jacopo Marconi}
\affiliation{ 
Dipartimento di Meccanica, Politecnico di Milano, Milano, Via La Masa 1, 20156, Italy
}
\author{Alper Erturk}%\altaffiliation[Also at ]{Georgia Institute of Technology, George W. Woodruff School of Mechanical 
\affiliation{
George W. Woodruff School of Mechanical Engineering, Georgia Institute of Technology, Atlanta, 30332, USA
}
\author{Massimo Ruzzene}%\altaffiliation[Also at ]{Georgia Institute of Technology, George W. Woodruff School of Mechanical 
\affiliation{ 
Daniel Guggenheim School of Aerospace Engineering, Georgia Institute of Technology, Atlanta, 30332, USA
}
\affiliation{
George W. Woodruff School of Mechanical Engineering, Georgia Institute of Technology, Atlanta, 30332, USA
}

\date{\today}% It is always \today, today,

\begin{abstract}
We report on broadband-to-narrowband elastic wave filtering resulting from time-periodic modulation of the stiffness of a one-dimensional elastic waveguide. Time modulation produces flat dispersion bands at frequencies that are multiple integers of half the modulation frequency. These flat bands lead to the selective reflection of a broadband incident wave at the interface between a non-modulated medium, and one with time-modulated stiffness properties. This results from the vanishing group velocity at the flat band frequencies, which prevents their propagation into the modulated domain. Thus, the considered modulated waveguide is understood as a single port system, in which a broadband incident wave (input) results in a narrowband reflected wave (output) at a frequency defined by modulation. The appearance of the flat bands for a time-modulated waveguide is here illustrated analytically and through numerical simulations. The filtering characteristics of a non-modulate/modulated interface are observed experimentally by implementing a square-wave modulation scheme that employs an array of piezoelectric patches bonded to an elastic waveguide subject to transverse motion. The patches are shunted through a negative electrical capaticance that, when connected, implements a stiffness reduction for the resulting electromechanical waveguide. Switching the capacitance on and off effectively modulates the stiffness of the waveguide, and illustrates the filtering characteristics associated with time-modulation of the equivalent elastic properties. We envision that a similar approach could be extended to investigate other properties of time-modulated elastic metamaterials, such as non-reciprocity and one-way filtering of elastic waves.

%
%Valid PACS numbers may be entered using the \verb+\pacs{#1}+ command.
\end{abstract}

%Use showkeys class option if keyword
\maketitle

%\tableofcontents

%\section{\label{sec:level1}Introduction}
Time-dependent material properties have been the object of considerable attention over the years. Parametric effects in time-modulated media have long been used for amplification of electromagnetic waves ~\cite{Cullen1958,Tien} and surface acoustic waves~\cite{Chao1970,Smith1976}. Non-reciprocal elements based on both up and down converter amplifiers have been introduced in the early 1960s~\cite{Kamal1960}. The interest in time-modulated media, motivated by their application for parametric amplification in electromagnetic waveguides and for signal processing applications, has led to numerous studies of both periodic~\cite{Simon,Hessel1961,Cassedy1963,Holberg1966,Cassedy1967,Felsen1970} and non-periodic modulation schemes~\cite{Auld1968}. Recently, time-modulation of relevant physical properties, imposed in a traveling-wave form, has been explored to achieve non-reciprocal behavior not only in optics, but also in acoustics and mechanics~\cite{Sounas2017,PhysRevB.96.104110,TrainitiNJP2016}. For example, magnetless, efficient and compact radiofrequency communication systems are designed with spatiotemporally modulated gratings to be shielded from echos and reflections during transmission~\cite{Hadad29032016}. Similarly, cavities with time-dependent volume in subwavelength acoustic circulators~\cite{PhysRevB.91.174306} have shown to provide isolation levels as high as 40 dB within the audible range, while asymmetric transmission has been reported in an acoustic waveguide with a time-dependent scattering element~\citep{Wang2015}.  Furthermore, non-reciprocal components connected in 1D and 2D lattice arrangements~\cite{VILA2017363} have been investigated in the pursuit of non-trivial wave topologies characterized by defect and backscattering immune propagation~\cite{Swinteck2015,Fleury2016}.

Time-modulation of electromagnetic waves can be achieved by modulating the material's permeability and permittivity simply by inducing magnetic and electric dipole moments~\cite{Lurie2007}. In mechanics, numerous studies have focused on the theoretical aspects of time-dependent material properties and their potential to produce non-reciprocity~\cite{NASSAR201710}. However, the practical implementation of dynamically changing stiffness or mass distributions remains an open challenge. Among the suggested approaches, light induced softening in $\mathrm{Ge_{x}Se_{1-x}}$ glasses has been explored in~\citep{PhysRevLett.92.245501}, while Coriolis-type effects have been exploited to produce a time-dependent moment of inertia in a pendulum with a radially moving mass~\cite{BELLINO2014120}. More recently, a phononic crystal with spatio-temporal modulation of electrical boundary conditions in a stack of piezoelectric elements has been described in~\cite{Croenne2017}. Magnetoelastic media interacting with an external magnetic field~\cite{Ansari2017} and magnetorheological fluids~\cite{Nanda2017} are also solutions recently suggested to implement traveling wave modulation.

In this Letter, we report on elastic waveguides whereby time-modulated stiffness produces frequency flat bands. We show that these bands can be exploited for the frequency-selective reflection of a broadband incident plane wave propagating in an unmodulated medium, which is interfaced with a time modulated one. The system is understood as a single port system, in which a broadband incident wave (input) results into a narrowband reflected wave (output) whose frequency content is centered at integer multiples of half the modulation frequency $\omega_m$. At the frequency corresponding to a flat band, the group velocity goes to zero, and propagation does not occur. As a result, these frequency components are reflected into the homogeneous domain. The location of the flat bands in the dispersion diagram is defined by the modulation frequency, which is a control parameter that may be tuned. The generation of the flat bands is first illustrated analytically for the case of a one-dimensional (1D) waveguide subject to transverse motion. Numerical simulations of a modulated medium interfaced with a non-modulated one illustrate the frequency selective reflection and therefore the filtering behavior of the system. The experimental implementation of the concept consists in an aluminum beam partially covered by an array of piezoelectric patches. When shunted through a negative capacitance circuit, the patches effectively reduce the equivalent stiffness of the beam~\citep{Marneffe2008}. Periodic operation of the switch that controls the connection of the patch with the negative capacitance produces square-wave modulation of the stiffness. Experimental measurements of wave propagation and reflection confirm the ability of the piezoelectric shunts to control the stiffness properties of the beam, and of the modulation to induce the narrowband filtering of the reflected wave at the interface between plain and stiffness modulated beam.

We first present an analytical model of an elastic, time-modulated waveguide, showing the existence of flat bands and their connection to the stiffness modulation frequency $\omega_m$. We consider the transverse motion of a beam with time-dependent material properties, which is governed by the following equation:
\begin{equation} \label{eq:eqmot2Beam}
R^2_g E(t)\frac{\partial^4 w(x,t)}{\partial x^4}+\frac{\partial}{\partial t}\bigg[\rho(t) \frac{\partial w (x,t)}{\partial t}\bigg]=0,
\end{equation}
where $A$, $I$  and $R_g=\sqrt{I/A}$ respectively denote the area, second moment and radius of gyration of the beam cross section, while $E$ and $\rho$ are the material Young's modulus and density. We introduce a time-dependent stiffness by assuming that the Young's modulus varies periodically in time, i.e. $E(t)=E(t+T_m)$, where $T_m=2 \pi /\omega_m$. A solution of the resulting equation of motion is sought in the form:
\begin{equation} 
w(x,t)=e^{i(\omega t - \kappa x)} \sum\limits_{n=-\infty}^{+\infty} \hat{w}_n e^{in\omega_m t }.\label{eq:sol}
\end{equation}
For simplicity of the derivations to follow, we assume an harmonic modulation, i.e. 
\begin{equation}
E(t) = E_0 + E_m \cos(\omega_m t) = E_0(1+\alpha_m \cos(\omega_m t) )
\end{equation}
where $\alpha_m=E_m/E_0$ defines the amplitude of modulation.

\begin{figure}
\includegraphics[width=0.8\textwidth]{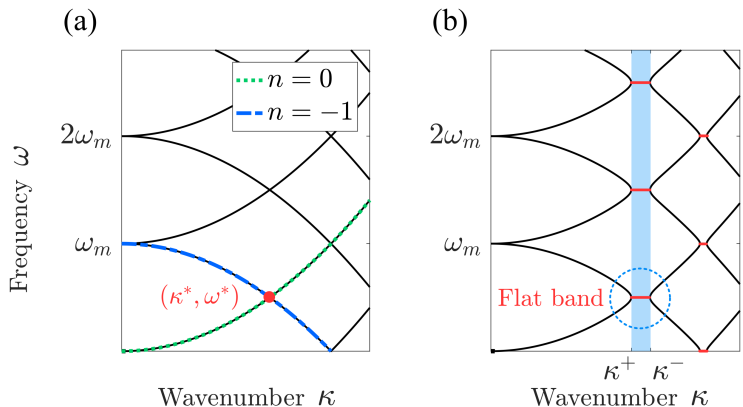}% Here is how to import EPS art
\caption{\label{fig:Fig1_kappa_bigger} Dispersion diagrams for a beam in bending: homogeneous case (a), time-modulated with harmonic modulation (b). The interaction between the $n=0$ and $n=-1$ order dispersion branches induces a flat band, in red, in a modulated system. Such flat band collapses to the point $(\kappa^*,\omega^*)$ for vanishing modulation amplitude. The approximate expression for the flat band, ranging from $\kappa^{+}$ to $\kappa^{-}$, is represented by the blue band.}
\end{figure}

The dispersion relation for the system is obtained by solving a quadratic eigenvalue problem (QEP) in $\omega$ upon imposing a wavenumber $\kappa$. The resulting dispersion diagrams for $\alpha_m\rightarrow 0$ and $\alpha_m\neq 0$ shown in Fig.~\ref{fig:Fig1_kappa_bigger} are $\omega_m$-periodic in the frequency domain, and notably feature flat bands for finite $\alpha_m$ values. These were previously observed and denoted as $\kappa$ bandgaps in photonic crystals with time-periodic dielectric properties~\cite{Zurita2009}. In analogy with frequency bandgaps, or stop bands, these flat bands in the wavenumber domain are associated with a stationary oscillating component corresponding to a zero group velocity, and an exponentially growing component  associated with parametric amplification. This is illustrated by imposing real values of wavenumber $\kappa$, and solving for the resulting complex frequency $\omega$.
Considering, for example, the $n=0$ and $n=-1$ orders only in Eq.~\ref{eq:sol}, leads to the following characteristic equation:
\begin{equation}\label{eq:CharactEq}
(\omega^2 - \gamma \kappa^4)\Big[ (\omega - \omega_m)^2 - \gamma \kappa^4 \Big] - \Big(\frac{\alpha_m \gamma \kappa^4}{2} \Big)^2 =0,
\end{equation}
with $\gamma=c_0^2 R_g^2$ and $c_0=\sqrt{E_0/\rho}$. For $\alpha_m\to0$, the $n=0$ and $n=-1$ order branches intersect at the point $\kappa^*=\sqrt{\omega_m/(2 c_0 R_g)}$ and $\omega^*=\omega_m/2$. For the modulated systems ($\alpha_m\neq0$), the solution of Eq.~\ref{eq:CharactEq} is instead:
\begin{equation}
\omega = \frac{1}{2}\Bigg[\omega_m - \sqrt{\omega_m^2 + 4\gamma \kappa^4 - 2 \sqrt{\gamma \kappa^4 (4 \omega_m^2 + \alpha_m^2 \gamma \kappa^4)}}    \Bigg].
\end{equation}

We observe that $\omega$ is complex if:
\begin{equation}
\omega_m^2 + 4\gamma \kappa^4 - 2 \sqrt{\gamma \kappa^4 (4 \omega_m^2 + \alpha_m^2 \gamma \kappa^4)}<0,
\end{equation}
which occurs for wavenumber values $\kappa \in [\kappa^{+},\,\,\, \kappa^{-}]$, with:
\begin{equation} \label{eq:kappaomegaim}
\kappa^{\pm}=\sqrt[4]{\frac{\omega_m^2}{2\gamma(2\pm\alpha_m)}} %, \quad \kappa^{-}=\sqrt[4]{\frac{\omega_m^2}{2\gamma(2-\alpha_m)}}.
\end{equation}
%with $\kappa^{+}=\sqrt[4]{\omega_m^2/[2\gamma(2+\alpha_m)]}$
In this wavenumber range, the corresponding frequency is complex, i.e. $\omega=\omega_r+i\omega_i$, where:
\begin{align}\label{eq:ori}
 \omega_r =& \frac{1}{2}\omega_m, \\ \omega_i =& -\frac{1}{2}\sqrt{\omega_m^2 + 4\gamma \kappa^4 - 2 \sqrt{\gamma \kappa^4 (4 \omega_m^2 + \alpha_m^2 \gamma \kappa^4)}}.
\end{align}
The real part of the frequency thus remains constant with respect to the wavenumber, which suggests a stationary wave at a fixed frequency as described by a zero group velocity, i.e.:
\begin{equation}
\frac{\partial \omega_r}{\partial \kappa}=0.
\end{equation}
Furthermore, we note that $\omega_i<0$, hence waves with wavenumbers satisfying Eq.~\ref{eq:kappaomegaim} are associated to an exponentially growing response in time. 

%Finally, for finite modulation amplitude, the $n=0$ dispersion branch and the higher-order dispersion branches interact, and this interaction results in flat bands, since:
%\begin{equation}
%\sqrt[4]{\frac{\omega_m^2}{2\gamma(2+\alpha_m)}}\leq\lim_{\alpha_m\to0} \kappa \leq \sqrt[4]{\frac{\omega_m^2}{2\gamma(2-\alpha_m)}}
%\end{equation}
%leads to:
%\begin{equation}
%\kappa^* \leq \kappa \leq  \kappa^* \quad \to \quad \kappa\equiv\kappa^*.
%\end{equation}
We remark that the solution of the QEP obtained by imposing the solution in eq.~\ref{eq:sol} leads to a family of branches that are identified by integers $n$. From the solution of the characteristic equation for various index orders and for $\alpha_m=0$ one can show that these branches intersect at frequencies $\frac{|n-n'|}{2} \omega_m$, where $n$ and $n'
$ denote two intersecting branches. These frequency values also identify the flat branches that occur when $\alpha_m\neq0$.

%here that, due to the periodicity of the dispersion diagram in the frequency domain, the same conclusion drawn so far can be extended to all the other flat bands at frequencies $\omega_m/2 + p \omega$, with $p$ integer. Furthermore, we notice that similar arguments could be extended to flat bands arising from the interaction between the $n=0$ dispersion branch and any other $|n|>1$ dispersion branch.

\begin{figure}
\includegraphics[width=0.5\textwidth]{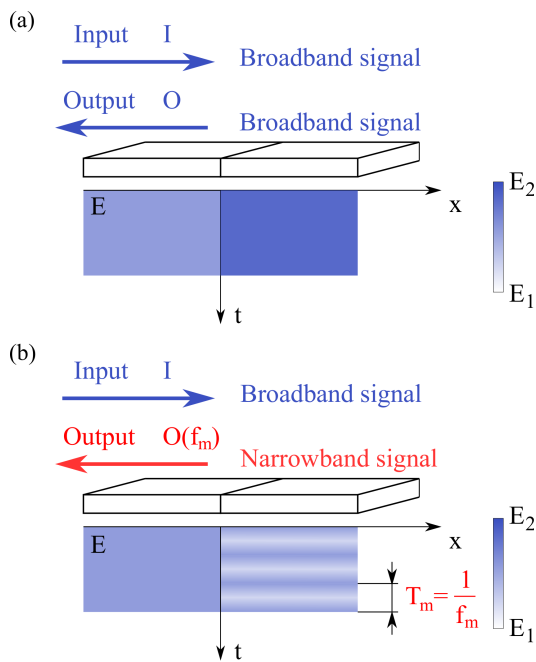}% Here is how to import EPS art
\caption{\label{fig:Fig2_cinque} 
Single port system converting a broadband signal input and a narrowband output through time-modulation. Assuming the elastic moduli $E_2>E_1$, a broadband wave impinging on the interface due to an abrupt change in Young's modulus in a piecewise homogeneous structure generates a broadband reflection (a). On the contrary, a narrowband output is induced by the same broadband signal input at the interface between an homogeneous and time-modulated domains (b), with the frequency of modulation $f_m=1/T_m$ that can be used to tailor the frequency content of the output.}
\end{figure}

\begin{figure*}
\includegraphics[width=0.99\textwidth]{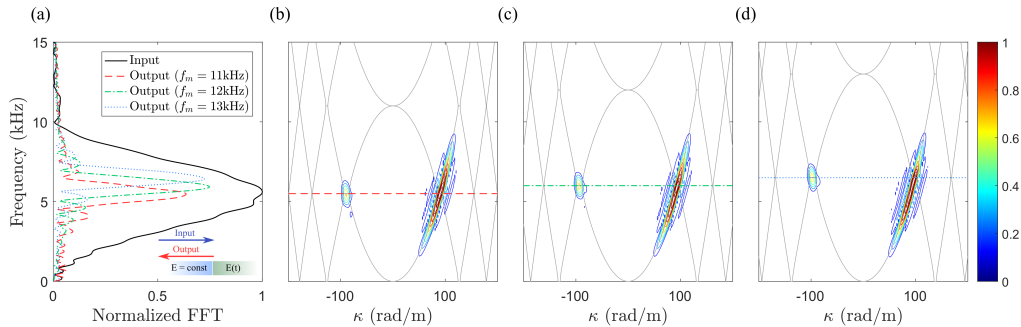}% Here is how to import EPS art
\caption{\label{fig:Fig3_edited} Response of the single port system obtained with a FDTD approach. For three different modulation frequencies, the same broadband input is converted into a different narrowband output (a). 
The Fourier transform's magnitude $|\mathcal{\hat{W}}(\kappa,\omega)|$ associated to the structure's response $w(x,t)$ shows narrowband frequency selection in reflected waves for broadband incident waves (b-d). Specifically, the reflected waves, which correspond to negative values of the wavenumber $\kappa$, have frequency content corresponding to the first flat band of the modulated structure at half of the frequency of modulation $f_m/2$, as predicted by the theoretical dispersion diagram (grey lines).}
\end{figure*}

Next, we investigate the behavior of a plane wave incident on an interface between a beam with constant properties and a modulated one. As discussed, we conjecture that wave components at frequency $\omega_m/2$, and its integer multiples, do not propagate through the modulated medium as a result of the occurrence of flat bands. Thus, a broadband wave incoming from the time constant domain will be transmitted with the exception of the flat band frequency, which is reflected. Interestingly, the same phenomenon could be explained in terms of power conversion phenomena between the different harmonic components at the interface, as done for waves propagating from homogeneous to time-modulated domains in dispersionless rods undergoing longitudinal motion~\cite{PhysRevB.96.104110}. Considering the broadband signal as the input and the narrowband signal as the output of a single port device, time modulation leads to the narrow band filtering of the input signal. The filtering frequency of such system, depicted in Fig.~\ref{fig:Fig2_cinque}, can be selected to produce different narrowband frequency outputs for the same broadband input through the selection of the modulation frequency $\omega_m$.
The concept is illustrated by evaluating the transient response of the waveguide with interface, which is computed through a finite-difference time-domain (FDTD) approach. In the simulations, we assume the time constant domain to be of length $L_{h}=0.3$ m, while the modulated one is $L_{m}=0.48$ m long. The cross-section of the beam is rectangular with $R_g= 8.67\times10^{-4}$ m. The density of the material is $\rho = 2700$ $\mathrm{kg/m^3}$ , while the Young's modulus is $E_0=69.9$ GPa. In the time modulated domain, the Young's Modulus obeys a square modulation law:
\begin{equation}\label{Eq:SM}
E(t) = E_0 + \frac{\alpha_m E_0}{2}\bigg\{ \sgn\Big[\cos(\omega_m t)\Big]-1\bigg\},
\end{equation}
with $\alpha_m=0.2$. A wave is injected in the system through a perturbation applied at the free end of the time constant beam which varies in time as a 4-cycle tone-burst of center frequency $f_{ext} = 5$ kHz. %Results are computed for three modulation frequencies $f_m = 11, 12, 13$ kHz.
The frequency content of input $w_{in}(x_p,t)$ and output $w_{out}(x_p,t)$ are evaluated by probing a single location $x_p$ close to the interface in the time constant domain. The corresponding Fourier transform shown in Fig.~\ref{fig:Fig3_edited}.a displays the frequency bandwidth of the input signal (black solid line). A comprehensive representation of the wave motion $w(x,t)$ along the waveguide is obtained through its representation in the frequency/wavenumber domain $\mathcal{\hat{W}}(\kappa,\omega)$, which is obtained through the application of a Fourier transformation in space and time. The contour plots in Fig.~\ref{fig:Fig3_edited}.b-d correspond to the magnitude $|\mathcal{\hat{W}}(\kappa,\omega)|$ of the resulting quantity and effectively locate the spectral content of the wave field along the theoretical dispersion branches (gray lines), which are superimposed for convenience. The contours in the $\kappa>0$ region correspond to right propagating waves, while those in the $\kappa<0$ half plane are associated with left traveling, hence reflected wave components. This representation hence effectively illustrates how broadband incident (right propagating) signal encounters at the interface is reflected (left propagating)  as a narrowband signal at the frequency associated with half of the considered modulation frequencies, in this case chosen as $f_m = 11, 12, 13$ kHz (see Fig.~\ref{fig:Fig3_edited}).

\begin{figure}
\includegraphics[width=0.45\textwidth]{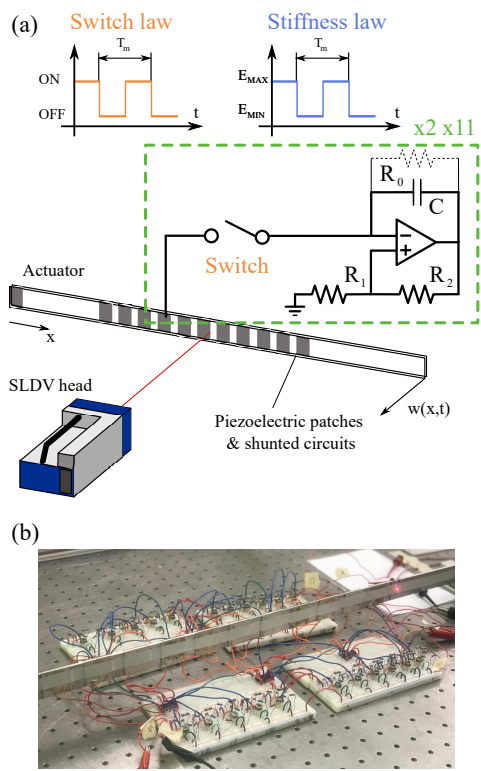}% Here is how to import EPS art
\caption{\label{fig:Fig4} Experimental validation of time-modulation of the stiffness properties in a beam through negative capacitance shunts and switches. The beam is equipped with 11 pairs of piezoelectric patches, each connected to a negative capacitance (NC) circuit (a). A switch opens and closes the patch-NC circuit series with a periodic law, inducing the stiffness to vary between two values with period $\omega_m$ (b). The response $w(x,t)$ is measured by a scanning laser Doppler vibrometer (SLDV).}
\end{figure}

%\subsection{\label{sec:level23}Experimental validation}
We achieve time-modulation of the stiffness of the considered elastic waveguide by employing an array of piezoelectric transducers bonded to the beam and shunted through a suitable electrical impedance. The resulting electro-mechanical waveguide is known to have an effective modulus of elasticity that depends on the electric impedance of the shunting circuit~\cite{Forward:79,HAGOOD1991243}, as result of the strain-voltage coupling inherent to the piezoelectric effect. %In the case of transverse motion of an elastic beam for example, the effect of the piezoelectric transducer on the stiffness can be described as a voltage-induced bending moment, in turn generated by the strain applied to the patch~\cite{Fuller1996}. The actuation-induced bending moment can favor or oppose the motion, thus reducing or increasing the structure's effective stiffness. However, the use of the piezoelectric transducers alone marginally affects the dynamics of the structure, therefore they are often coupled to electrical circuits.
For example, resonant moduli of elasticity have been exploited to induce tunable band gaps at frequencies defined by the resonant characteristics of the shunts~\cite{Airoldi2011}, while broadband stiffness modulus control has been achieved through negative capacitance (NC) circuits~\cite{Behrens2003}. According to ~\cite{Behrens2003,Beck2013}, the elastic modulus $E_p^{SU}$ of a piezoelectric patch connected to a NC $-C'$ can be expressed as:
\begin{equation}\label{Eq:neg capacitacne}
E_p^{SU}=E_p^E \frac{C'-C_p^T}{C'-C_p^S}
\end{equation}
where $E_p^E$ is the elastic modulus of the piezoelectric patch with short-circuited electrodes,  $C_p^T$ and $C_p^S=C_p^T(1-k_{31}^2)$ respectively are the stress-free and strain-free piezoelectric capacitance values, and $k_{31}$ is the piezoelectric coupling coefficient corresponding to the longitudinal straining of a through-the-thickness polarized patch. Values of $C'>C_p^T$ ensure stability of the patch as an electromechanical system~\cite{Marneffe2008}, while producing significant changes in elastic modulus with respect to the open circuit value $E_p^D=E_p^E / (1-k_{31}^2)$, which is obtained for $|C'|\rightarrow 0$. The negative capacitance is implemented by considering the negative impedance converter circuit of Fig.~\ref{fig:Fig4}.a, which implements a capacitance $C_N=-C'=-R_2/R_1 C$~\cite{Behrens2003}. The additional resistor $R_0$ prevents saturation of the capacitor, which would lead to instability~\cite{Beck2013}. A square wave time-modulation is obtained by operating a switch that breaks the series connection between the piezoelectric transducer and the NC shunt, thus inducing the elastic modulus of the piezoelectric patch to vary between the closed circuit ($E_p^{SU}$) and the open circuit ($E_p^D$) values (see eq.~\eqref{Eq:neg capacitacne}).  

The experimental set-up employs a slender aluminum beam ($E=69.9$ GPa, $\rho = 2700$ $\mathrm{kg/m^3}$) with  $A= h\times b$ rectangular cross section ($h=3$ mm, $b=3$ cm), and an array of 11 pairs of piezoelectric patches (Young's modulus $E_p^E=73$ GPa, density $\rho_p=7800$ kg/m$^3$, thickness $h_p=1$ mm and width $b_p=3$ cm) vacuum-bonded to the beam at regular spatial intervals separated by a $1.5$ cm gap.  (Fig.~\ref{fig:Fig4}). All patches all connected in series to a NC shunt with negative capacitance $C_N=-C'=-11$ nF which approaches the stress-free capacitance of the piezo patches, which is equal to $C_p^T=9.06$ nF. This value corresponds to a Young's modulus reduction $E_p^{SU}/E_p^{D}\approx 60\%$. The overall effect on the bending stiffness of the beam is estimated to be approximately equal to a $14\%$ reduction, which accounts for the stiffness corresponding to the aluminum base beam, and the patches based on the considered filling ratio. Details on the estimation of the equivalent stiffness change are provided in the Supplementary Material.

The switches between the piezoelectric transducers and the NC shunts are operated at a modulation period $T_m$, which induces a square wave modulation of fundamental frequency $\omega_m=2 \pi/T_m$. The beam is excited by a piezoelectric transducer bonded close to its free end (See Fig.~\ref{fig:Fig4}), which induces a transversely polarized wave propagating along the beam length. The corresponding velocity field is measured by a scanning laser Doppler vibrometer (SLDV), also shown in Fig.~\ref{fig:Fig4}, along a line of 
$512$ equally spaced points that covers the entire length of the beam. Additional detail on data acquisition and experimental parameters are provided in the Supplementary Material. The recorded spatio-temporal wavefield $w(x,t)$ is analyzed in the frequency/wavenumber domain, where incident and reflected components are separated. This allows the identification of input $w^{(in)}(x_p,t)$ and output $w^{(o)}(x_p,t)$ components at the probe location $x_p$, which as in the numerical investigation, is near the interface between homogeneous and time-modulated domains.
	
\begin{figure*}
\includegraphics[width=0.99\textwidth]{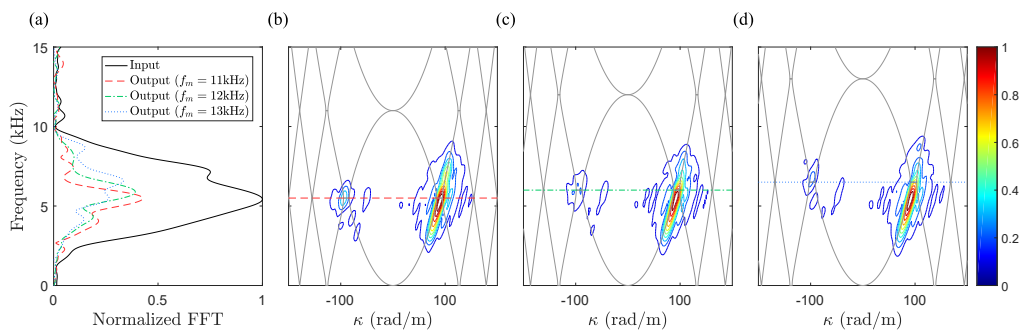}% Here is how to import EPS art
\caption{\label{fig:Fig5_edited} 
Measurement of the wave filtering properties in time-periodic beam with SLDV. The frequency spectrum of the reflected waves is centered at $f_m/2$, as expected for all three measurements (a). The Fourier transform's magnitude $|\mathcal{\hat{W}}^{exp}(\kappa,\omega)|$ of the structure's response $w^{exp}(x,t)$ in the modulated domain shows that the reflected wave depends on the modulation frequency, as predicted by the flat bands in the theoretical dispersion diagram (grey lines).}
\end{figure*}

Three experiments are performed by inducing a broadband excitation signal, centered at 5 kHz, and modulating the effective stiffness at frequency $f_m =$ 11, 12 and 13 kHz. The analysis of the frequency spectrum of incident and reflected wave components at the probe location (Fig.~\ref{fig:Fig5_edited}.a) confirms that the reflected waves are narrowband at a frequency $f_m/2$ defined by the applied modulation. This is also illustrated in the contour plots of Fig.~\ref{fig:Fig5_edited}.b-d which display the frequency/wavenumber content of the measured wavefield for the considered values of the modulation frequency. While the dispersion branch associated with the incident wave in the $\kappa>0$ region remains unaltered in all the three experiments, the energy of the reflected wave, located in the $\kappa<0$ region, is narrowband. We conclude that the system indeed behaves as a port, transforming the incident broadband signal into a narrowband signal. Furthermore, the location of the energy content of the output is corresponds to the flat bands at $f_m/2$, in our experiments at 5.5, 6 and 6.5 kHz, in accordance to our simple theoretical model.

%The experimental results in Fig.~\ref{fig:Fig5_edited} very well confirm the theoretical predictions of our simplified model. We considered the three modulation frequencies already assumed for the numerical simulations, \emph{i.e.} $f_m$ equal to 11, 12 and 13 kHz. Clearly, the time-modulation of the beam's stiffness results in reflected waves with narrowband frequency content centered at $f_m/2$, which corresponds to the first flat band of the time-modulated structure.

In conclusion, we have experimentally implemented a broadband-to-narrowband single-port filter of elastic waves based on stiffness time-modulation. The system is implemented in the form of a beam in transverse motion, endowed with an array of piezoelectric patches bonded to it and connected to switchable negative capacitance shunt. Softening is induced in each piezoelectric patch when the connected in series with the negative capacitance shunt. A switch periodically breaks the series connection between the transducer and the shunt, making the effective stiffness of the structure vary with time at a certain modulation frequency $\omega_m$. A simplified analytical model of the system predicts the existence of wavenumber stop bands, or flat bands, at integer multiples of $\omega_m/2$. The flat bands predict the narrowband frequency content of the reflected waves, the output of the port, given a broadband signal in input. Importantly, as shown both by numerical simulation and experiments, the single-port system can be tuned at will by changing the modulation frequency $\omega_m$, thus dramatically affecting the frequency content of the output. 

\begin{acknowledgments}
The work is supported by the National Science Foundation - CMMI/ENG Division, through grant 1719728.
\end{acknowledgments}

\bibliography{apssamp}% Produces the bibliography via BibTeX.

\end{document}